# Excitation spectrum of a model antiferromagnetic spin-trimer

M. B. Stone

*Neutron Scattering Science Division, Oak Ridge National Laboratory, Oak Ridge, Tennessee 37831, USA*

F. Fernandez-Alonso

*ISIS Facility, Rutherford Appleton Laboratory, Chilton, Didcot, Oxfordshire OX11 0QX, United Kingdom*

D. T. Adroja

*ISIS Facility, Rutherford Appleton Laboratory, Chilton, Didcot, Oxfordshire OX11 0QX, United Kingdom*

N. S. Dalal

*Department of Chemistry and Biochemistry Center for Magnetic Resonance, National High Magnetic Field Laboratory, Florida State University Tallahassee, FL 32306, USA*

D. Villagrán

*Department of Chemistry, Texas A & M University, College Station, TX 77843, USA*

F. A. Cotton

*Department of Chemistry, Texas A & M University, College Station, TX 77843, USA*

S. E. Nagler

*Neutron Scattering Science Division, Oak Ridge National Laboratory, Oak Ridge, Tennessee 37831, USA*

**ABSTRACT**

We present an inelastic neutron scattering (INS) study of the excitation spectrum of a model $S = 1/2$ equilateral Heisenberg trimer, $Cu_3(O_2C_{16}H_{23})_6 1.2C_6H_{12}$. Earlier measurements were interpreted on the basis of a spin ground state consisting of a pair of degenerate $S = 1/2$ doublets, with an $S = 3/2$ quartet excited state. Given that the simplest model of magnetism for this compound includes a degenerate ground state, using thermodynamic probes to determine the excitation energy between the ground and excited states is challenging. An estimate of the excitation energy between the ground and excited states based on magnetic susceptibility measurements is roughly ~28 meV. INS measurements of this compound are likewise challenging since inter- and intramolecular vibrational modes associated with the organic ligands are at frequencies similar to the magnetic excitations. However, by measuring a non-magnetic analog, *i.e.* ligands only, as well as the temperature dependence of the excitations in the magnetic system, we are able to determine the excitation energy as being approximately 36.5 meV at $T = 10$ K, with a gradual softening to 32 meV with increasing temperature. We emphasize the consideration of ligand/lattice based excitations while interpreting magnetic excitations of such nanomagnetic systems containing organic molecules while underscoring the importance of INS in determining the magnetic excitation spectrum in systems with degenerate ground states.

Inelastic neutron scattering (INS) is a powerful probe of molecular magnets and small magnetic clusters because of the ability to directly couple to the magnetic excitation spectrum of these systems. A number of such materials have been examined with neutron scattering techniques. Typically these have been based upon large clusters of $S > 1/2$ magnetic ions as measured in manganese[1,2,3], iron[4,5,6], nickel[7,8] or chromium[9,10] spin-clusters. The analysis of spectroscopic data from these systems is found to be complex because of the high degree of spin frustration and multiple exchange constants within the molecule[11]. For example, neutron[4] as well electron paramagnetic resonance measurements on $Fe_8$[12] clearly suggest that the $S = 10$ ground state is nested with at least the $S = 9$ excited state and possibly others. Similar observations of a series of closely spaced excitations in the vicinity of the ground state have also been reported for $Mn_{12}$-acetate[13,14,15]. The large number of magnetic interactions in such magnetic clusters often results in a series of degenerate or near degenerate ground states analogous to those found in highly frustrated three-dimensional magnetic systems such as the 'spin-ice' pyrochlores $Dy_2Ti_2O_7$[16], $Ho_2Ti_2O_7$[17] and $Ho_2Sn_2O_7$[18]. There have also been numerous studies of systems composed of small clusters of spins including dilute magnetic insulators [19,20], as well as spin-dimers, trimers and tetramers residing in both inorganic [21,22,23] and organic [24,25,26,27] matrices. The isolated nature of these spin-clusters and their limited number of magnetic sites allows for the calculation of spin-excitations, thermodynamics and scattering cross-sections to be more tractable[28,29]. In contrast, many of the large spin-cluster compounds of interest are synthesized as metal clusters in large organic groups (*e.g.* phenyl or other functional groups) that tend to complicate measured INS spectra. The compound we examine is similarly constructed, but with a much simpler spin-cluster composed of an equilateral trimer of quantum spins. The antiferromagnetic spin-trimer excitation spectrum is well understood in principle and serves as one of the simplest examples of geometric frustration, illustrating the complications associated with degenerate ground states[30,28,31]. Furthermore, the rotational and permutational symmetry of the equilateral Heisenberg trimer make its excitation spectrum unique compared to the general spin-trimer or isosceles cases.

We examine the excitation spectrum of the trinuclear compound $Cu_3(O_2C_{16}H_{23})_6 \cdot 1.2C_6H_{12}$, hereafter referred to as $Cu_3$, where the magnetic species are quantum spin, $S = 1/2$, $Cu^{2+}$ ions. The original structure determination via x-ray scattering shows that the $Cu^{2+}$ ions reside on the vertices of an equilateral triangle, as shown in Fig. 1(a), with six surrounding 2-4-6 triisopropyl-benzoic acid (HTiPB) carboxylate ligands, as shown in Fig. 1(b)[32]. The trimer can be described by the discrete spin-Hamiltonian $H = J(S_1 \cdot S_2 + S_2 \cdot S_3 + S_3 \cdot S_1)$, where $S_n$ are the individual $S = 1/2$ sites and $J$ is the magnitude of the antiferromagnetic exchange between spins. The eigenvalues of this Hamiltonian consist of a degenerate ground state pair of $S_{Total} = 1/2$ doublets separated by $\Delta = 3/2J$ from a $S_{Total} = 3/2$ quartet excited state as shown in Fig. 2(a) case(i). This contrasts the lower symmetry isosceles case which has a single doublet ground state with two transitions out of the ground state as shown in Fig. 2(a) case(ii). In an applied magnetic field, the degeneracy of these levels is further lifted as shown in case(iii) of Fig. 2(a). Bulk magnetization measurements as a function of temperature agree well with the equilateral triangle description for $Cu_3$, and indicate that the three spins are coupled antiferromagnetically with isotropic Heisenberg exchange $J = 18.6$ meV[32,33]. EPR spectroscopic measurements are also consistent with this description of the Cu$_3$ compound, although at low-temperatures, $T \leq 10$ K, there appears to be a very small distortion away from the equilateral arrangement of spins and/or a greater influence of additional inter-trimer exchange parameters of the order of $J' = 0.15$ K[34]. The measurements we perform here however are insensitive to this small effect, $J/J' \approx 7.21$E-4. Thus it seems that Cu$_3$ is essentially expected to be a clean model equilateral trimer Heisenberg antiferromagnet.

The INS cross-section based upon the isolated $S = 1/2$ trimer model consists of a single non-dispersive excitation at energy transfer $\hbar\omega = \Delta = 3/2J$. The powder averaged wave-vector, $Q$, dependent scattering intensity of the $S = 1/2$ trimer is calculated as

$$I(Q) \propto |F(Q)|^2 \left[1 - \frac{\sin(Qd)}{Qd}\right] \delta(\hbar\omega - \frac{3}{2}J), \qquad (1)$$

where $d$ is the distance between $S = 1/2$ sites and $F(Q)$ is the magnetic ion form factor. Any inter-trimer coupling would serve to broaden the energy of the single-mode in a powder measurement. We show Eq. (1) and its individual components evaluated for the specific parameters of the Cu$_3$ compound in Fig. 2(b). The product of the magnetic form factor and the trimer structure factor results in an asymmetrically peaked function with a maximum at $Q = 1.29$ Å$^{-1}$ and a less defined shoulder in the vicinity of $Q = 3.4$ Å$^{-1}$.

Although the description of magnetic excitations and the INS cross section of the Heisenberg equilateral trimer are straightforward, one must be cognizant of the fact that molecular vibrations associated with the organic portions of the system may be at frequencies very similar to the magnetic excitation spectrum. Because of this difficulty, we also examine a non-magnetic analog sample to distinguish magnetic and vibrational modes in a technique that has been applied successfully in the study of heavy fermion systems with neutron spectroscopy[35,36]. We examine the molecular excitations directly through measurements of the 2-4-6 triisopropyl-benzoic acid (HTiPB) organic ligands. Such measurements will not probe the identical acoustic phonon modes as in the $Cu_3$ compound, but the molecular vibrations associated with HTiPB and the HTiPB portions of the $Cu_3$ compound will be at very similar frequencies. From inspection of the crystal structure, there is also clear potential for large magnon-phonon coupling in the $Cu_3$ system. This is due to the fact that small changes in bond distances and angles of the compact magnetic core of the molecule shown in Fig. 1(a) will significantly affect the exchange parameters of the trimer. Magnon-phonon or more specifically magnon-libron coupling would result in either a repulsion of the respective modes, or an overall broadening of the magnetic excitation[37,38]. As we discuss below, in examining our INS measurements to yield a measure of the excitation energy in $Cu_3$ it is extremely important to examine vibrational modes in the excitation spectra.

EXPERIMENTAL TECHNIQUES

Powder samples of $Cu_3$ were synthesized from 2-4-6 triisopropyl-benzoic acid $[C_{16}H_{24}O_2$ (HTiPB)] and copper carbonate $[CuCO_3 \cdot Cu(OH)_2]$ as discussed elsewhere[32]. The resulting dark-green solid was dissolved in cyclohexane, the solution was then volume-reduced and stored at 10 °C. Well-formed dark green crystallites appear after about a week. The composition of these crystals is $Cu_3(TiPB)_6 \cdot 1.2 C_6H_{12}$. The crystal structure has been reported in detail elsewhere[32]. The metallic core of this unit is shown in Fig. 1(a) and consists of three $Cu^{2+}$ ions in an equilateral triangle arrangement in the *ab* plane. The Cu-Cu linear distance is 3.131(3) Å, and the Cu sites are bonded to one another by two HTiPB carboxylate ligands, one of which is shown in Fig. 1(b).

INS measurements were performed on a 0.397 gram powder sample of Cu$_3$ (magnetic sample) and a 2.089 gram sample of HTiPB (non-magnetic sample consisting of ligands only) using the HET time-of-flight spectrometer[39] at the ISIS neutron scattering facility. Samples were sealed in aluminum sample holders in an argon atmosphere. The incident energy, $E_i$, used for individual measurements is noted in figures of the respective data. Different chopper frequencies were chosen for each $E_i$ in order to preserve energy resolution and maintain neutron flux incident on the sample. Several incident energies were chosen to provide spectra over different ranges of energy transfer, up to approximately 330 meV. Temperature control between $T = 5$ K and $T = 295$ K was provided by a top-loading closed cycle refrigerator.

**RESULTS AND DISCUSSION**

Figure 3 shows three low-temperature, $4 \leq T \leq 11$ K, and one mid-temperature, $T = 80$ K in Fig. 3(c), wave-vector integrated spectra obtained for Cu$_3$. Three of these measurements were also performed with the HTiPB sample. The EPR and bulk magnetization measurements[32,33,34] predict a single magnetic excitation at an energy transfer of $\hbar\omega = \Delta = 3/2J \approx 28$ meV. Figures 3(b) and 3(c) both indicate a broad intense peak centered at 33.7(2) meV, which as we show later is composed of two individual peaks in the vicinity of 31 meV and 37 meV. A comparison of the Cu$_3$ and HTiPB results indicates that the magnetic and non-magnetic sample share similar spectra in the vicinity of the proposed magnetic transition. The incoherent cross section associated with the protonated HTiPB ligands is responsible for contamination of the magnetic signal with a large background of lattice/molecular excitations.

Figures 3(b) and 3(c) indicate a broadening of the spectrum in the vicinity of 30 meV for the $Cu_3$ measurement compared to the measurements of the non-magnetic HTiPB sample. This is likely due to the presence of both a molecular excitation and a magnetic excitation at similar energies in the $Cu_3$ sample, whereas there is only the single libron mode in the HTiPB sample. We use the $E_i = 112$ meV measurements to characterize these excitations. Fitting the HTiPB data in Fig. 3(c) to a single inelastic Gaussian with an elastic Gaussian and sloping background, allows for a determination of the effective energy resolution of the inelastic excitation, full width at half maximum (FWHM) = 8.070(2) meV, with a peak location of 31.539(1) meV. Fitting the $Cu_3$ data to the same function yields a significantly broader peak, FWHM = 10.0(7) meV, centered at 33.5(2) meV, even though both measurements were performed under identical conditions and thus would share the identical resolution function. This indicates the potential for two neighboring peaks in the $Cu_3$ spectrum in the vicinity of 30 meV. Fixing the energy resolution to that found for the non-magnetic HTiPB sample but including two inelastic peaks in the fit of the $Cu_3$ data allows for a determination of these two excitations. This yields an identical lineshape to the single broad Gaussian peak with peaks located at 31.9(6) and 37(1) meV as shown in Fig. 3(c). The higher energy excitation in the $Cu_3$ measurement is likely the magnetic excitation of the trimer. In contrast, the much higher energy excitations shown in Fig. 3(d) for both the $Cu_3$ and HTiPB samples have nearly identical peak positions and widths as illustrated by multiple Gaussian fits of the two measurements ($Cu_3$: $\hbar\omega = 123(1)$, $\hbar\omega = 175.4(5)$ and $\hbar\omega = 216(2)$ meV with respective FWHM = 30(3), 29(1) and 13(3) meV) (HTiPB: $\hbar\omega = 121(1)$, $\hbar\omega = 175.4(4)$ and $\hbar\omega = 215(2)$ meV with respective FWHM = 30(3), 28(1) and 15(5) meV). Unlike the measurements shown in Fig. 3(c), these higher energy excitations are solely due to molecular vibrations associated with the organic ligand common to both samples.

Figure 2(b) shows that the scattering intensity predicted for the magnetic trimer excitations is a rapidly changing function of wave-vector transfer with a decaying scattering intensity at larger $Q$. Lattice/libron excitations however are typically characterized with a quadratic wave-vector dependent scattering intensity. Figure 4 depicts a series of constant wave-vector scans from the $E_i$ = 112 meV Cu$_3$ measurement. Similar to the treatment performed for the wave-vector integrated data discussed with respect to Fig. 3(c), these data along with additional scans at intermediate wave-vectors were simultaneously fit to a pair of inelastic Gaussians with a an elastic Gaussian peak and a sloping background. The FWHM of the inelastic Gaussians was held fixed to the value determined for the nonmagnetic sample, and the location of the inelastic Gaussians was a common fitting parameter. This resulted in values of 30.3(3) meV and 36.1(4) meV for the inelastic peak locations, quite similar to the values determined from the wave-vector integrated data in Fig. 3(c). We also perform this analysis for the $E_i$ = 61 meV measurements of Cu$_3$ shown in Fig. 3(b). This results in peaks located at energy transfers of 30.76(3) and 37.02(4) meV with fixed FWHM of 3.61(3) meV. The fitted series of constant wave-vector scans also yields the wave-vector dependent scattering intensity of these two peaks. Figure 5(a) and (b) show the fitted peak intensities as a function of wave-vector transfer for the ~36.5 and ~30.5 meV excitations as well as several other excitations as measured in both Cu$_3$ and HTiPB. All of the excitations, except the ~36.5 meV mode, in both the Cu3 and HTiPB samples agree well with the wave-vector dependence of a phonon/libron excitation. These data are fit to a quadratic function with a constant offset to account for multiple scattering due to the large hydrogen cross-section of the samples. Alternatively, the ~36.5 meV excitation of the $Cu_3$ sample does not follow the quadratic trend associated with a phononic/libronic excitation at wave-vectors in the vicinity of the rounded shoulder of the magnetic trimer scattering intensity shown in Fig. 2(b). This is further evidence that the magnetic trimer excitation in $Cu_3$ is in the vicinity of ~36.5 meV. Although the wave-vector dependence of this mode does not agree with the wave-vector dependent scattering intensity predicted in Fig. 2(b), we later resolve this disagreement by carefully examining the temperature dependent scattering intensity of magnetic and libron modes.

The magnetic excitation observed by INS, $\Delta$ =36.5 meV, contrasts the value of $\Delta$ = 28 meV inferred from measurements of temperature dependent magnetic susceptibility. This discrepancy is due to the difficulty in extracting a value of the excitation energy from thermodynamic measurements of systems with a degenerate magnetic ground state. When plotting the temperature dependent magnetization in terms of measured magnetic susceptibility rather than the effective magnetic moment (as used in Ref. 32), one immediately sees the difficulty in accurately distinguishing the magnetic response of the trimer from any weak paramagnetic background. This is due to a diverging magnetic susceptibility at low-temperatures associated with the degenerate magnetic ground state in the system. Prior to further examination of the magnetic transition, we consider the likely origin of the other measured excitations in both $Cu_3$ and HTiPB.

The spectra measured for both $Cu_3$ and HTiPB samples, Figs. 3(a)-(d), show virtually identical transitions for $\hbar\omega > 50$ meV as already illustrated for the data in Fig.

3(d). We note that the low-temperature, $T$ = 20 K, vibrational spectrum of several benzoic acid isotopomers has recently been measured via INS using the TOSCA spectrometer at ISIS [40,41]. The TOSCA spectrometer is an inverted geometry instrument and HET is a direct geometry instrument. Because of differences in energy resolution and recoil effects, comparison of peak position is not absolute but can only be made at a semi-quantitative level. Thus, the present set of measurements may be regarded as complementary to those from the TOSCA spectrometer. The TOSCA measurements identified many of the transitions of benzoic acid isotopomers through numerical calculations of the spectra. We provide spectroscopic labels for many features in our data in comparison to these higher resolution results. Both our $Cu_3$ and HTiPB samples show similar INS spectra up to approximately 250 meV. This corresponds to a region of single-bond stretch modes, *e.g.* CC, CO, as well as bending modes of lighter species, *e.g.* CH. From this observation, we conclude that our measured spectral features are mostly due to intramolecular normal modes of the organic HTiPB ligand.

We label our spectral features *A* through $K^{42}$ in Fig. 3 and make direct comparisons to the experimental and numerical work of Plazanet *et al.*[40,41] on benzoic acid isotopomers. The low-energy broad spectrum between 0 and ≈ 15 meV, *A* in Figs. 3(a) and (b), corresponds to a series of closely spaced lattice and librational modes as also observed in IR and Raman measurements [43,44]. The *B* and *C* peaks at ≈ 18.5 meV (153 cm$^{-1}$) and ≈ 23.5 meV (189 cm$^{-1}$) are consistent with out-of-plane carbon bending modes, where the plane is the benzene ring in the HTiPB structure as shown in Fig. 1(b). As shown earlier, the broad peak centered at ≈ 34 meV (274 cm$^{-1}$), is composed of multiple peaks which we label as *D* and *E* as well as the magnetic excitation in the $Cu_3$ sample.

This energy range agrees well with the designation of in-plane carbon bending modes. In addition, the *E* and *F* peaks correspond to in-plane carbon stretch and out-of-plane ring bending modes respectively.

We observe a small peak at ≈ 225 meV, *K*, corresponding to a CO stretch mode. The large intensity of the two spectral features at ≈ 120 and just below ≈ 180 meV, features *H* and *I*, imply that they involve the motion of hydrogen atoms. Spectral features due to the hydrogen sites will have an overwhelming amount of scattering intensity compared to other modes, due to its large incoherent scattering cross-section. The most likely candidates for these higher energy excitations are carbon-hydrogen in-plane and out-of-plane bending modes.

We caution that an initial analysis of our $Cu_3$ data, without comparison to the HTiPB sample, would reveal a spectral line in approximately the correct location, Fig. 3(c), as predicted based upon magnetic susceptibility and EPR measurements[32]. It would be misleading to interpret the temperature dependent scattering measurements of this feature, shown in Fig. 6, as solely magnetic in origin. The data in Fig. 6 indicate that there is a reduction and broadening in scattering intensity as temperature is increased. This is consistent with what is typically observed in magnetic systems; however, from our measurements of the non-magnetic HTiPB sample we know that there is a contribution of non-magnetic response to the scattering intensity in this region of energy transfer. For comparison, we examine the measured temperature dependence of three purely molecular excitations in the $Cu_3$ data at approximately 18, 45, and 55 meV, modes *B*, *F* and *G* respectively as shown in Fig. 3(b). The integrated intensities of the respective vibrational modes were first normalized by the ratio of the Bose occupation factor, $\langle n(\omega)+1 \rangle = (1-\exp[-\hbar\omega\beta])^{-1}$, where $\beta = \dfrac{1}{k_B T}$, for the particular mode to that of the proposed 36.5 meV magnetic excitation. The average temperature dependent relative normalized integrated intensity of these excitations is shown in Fig. 7(a). The temperature dependent normalized scattering intensity in the vicinity of 36.5 meV, integrated between 22 and 41 meV, for Cu$_3$ is plotted in Fig. 7(b). For typical organic vibrational modes the Debye-Waller factor implies that the scattering intensity will ultimately decrease as both wave-vector and temperature are increased[45,46,47,48]. The temperature dependent scattering intensity associated with such excitations is

$$Ip \propto \langle n(\omega)+1 \rangle \exp(-2W), \qquad (2)$$

where $\exp(-2W)$ is the temperature dependent Debye-Waller factor[49]. Because the vibrational modes and magnetic excitations are at similar energies, it is difficult to determine the appropriate temperature dependence of the Debye-Waller factor exactly. Furthermore, a complete determination of thermal parameters associated with the hydrogen sites has not been performed in the initial structural determination[32]. Rather we include the Debye-Waller factor as a fitting parameter of the form $2W = A_W$ Coth(½Δβ). The result of this single parameter fit to the data in Fig. 7(a), fixing $\varDelta = 36.5$ meV, yields a reduced $\chi^2 = 1.26$. Including other forms for $I_p$ with a temperature dependent vibration amplitude and other forms of the temperature dependent Debye-Waller factor yields similar results[45].

The temperature dependence of the magnetic scattering intensity, $I_m$, is[49]

$$I_m \propto (1+\exp[-\Delta\beta])^{-1}. \qquad (3)$$

We do not explicitly include a temperature dependent Debye-Waller factor for the magnetic excitations under the reasonable assumption that the compact tri-nuclear core of the $Cu_3$ molecule is much more rigid than the large surrounding organic groups. A comparison of Eq(3) to the data in Fig. 7(b) is not able to account for the rapid fall of in scattering intensity in the vicinity of 100 K. This is due to the fact that the low-temperature measurements are more significantly contaminated by vibrational mode scattering intensity than the high temperature measurements as illustrated in Fig. 7(a). Given that both vibrational and magnetic excitations are in the same vicinity, we model the temperature dependent scattering in Fig. 7(b) accounting for both contributions. We fit the relative integrated intensity to a sum of a magnetic term, Eq. 3, and a vibrational term, Eq. 2, fixing the Debye-Waller factor to the value found in analysis of the data in Fig. 7(a). Fixing the energy of the magnetic excitation to 36.5 meV, yields a single parameter fit for the ratio of vibrational scattering to magnetic scattering, which we label as $\rho$. The resulting fit is shown as a solid line in Fig. 7(b), and reproduces the measurement very well with reduced $\chi^2 = 0.926$. The fitted ratio of phonon scattering to magnetic scattering is found to be $\rho = 55(4)\%$. For comparison, we plot the individual magnetic and librational components of the normalized integrated scattering intensity in Fig. 7(b). From these components one can see that comparisons of the data in Fig. 7(b) to a purely vibrational (reduced $\chi^2 = 10.3$) or purely magnetic (reduced $\chi^2 = 56.8$) model are not appropriate.

We again note the potential for magnon-libron coupling in this system due to the large number of vibrational modes of the ligands which are connected to the compact magnetic trimer core of the molecule. Presumably, the tri-nuclear core is more rigid than the surrounding ligands shown in Fig. 1(b), such that magnon-libron effects would be due to vibrational modes of the carboxylate ligands interacting with the magnetic modes via the oxygen and carbon bonds linking the ligands to the magnetic trimer. We have noted that there is a broadening in the vicinity of the proposed magnetic excitation of the $Cu_3$ spectra in comparison to that of the organic ligands. Our temperature dependent measurements are also consistent with observing both intramolecular vibrations and magnetic excitations in this portion of the spectrum. From the temperature dependent INS measurements in Fig. 6 and the peak positions found in the analysis of the data shown in Figs. 3 and 4, we are able to quantify a temperature dependent softening of the mode which could be representative of magnon-libron coupling effects. Figure 8(a) shows the peak location of the proposed magnetic excitation as a function of temperature. The magnetic excitation energy associated with the data in Fig. 6 was determined from fitting two Gaussian peaks between 22 and 41 meV for $T < 150$ K and a single Gaussian peak at larger temperatures. There is a clear softening of the magnetic transition with increasing temperature. We compare these data to a smooth interpolating function of the form $\hbar\omega(T) = E_1 + (E_0 - E_1)\left(\dfrac{T^r}{T^r + T_0^r}\right)$, suggesting the low-temperature, $E_0 = 37(1)$ meV, and high-temperature, $E_1 = 31.1(5)$ meV, limits of the excitation energy. The value of $E_1$ is very similar to the fitted lower energy excitation shown in Figs. 3(c) and 4(a) which we associate with an in-plane carbon bending mode of the HTiPB ligand. Based upon our analysis and the crystal structure of the compound, there is clear potential for the involvement of mixed magnon-libron excitations, however a more detailed set of measurements would be required to draw further conclusions.

Magnon-libron interactions may result in a change in exchange energy, $J$, due to changes in crystal structure such as bond angles and distances. It is reasonable to expect a temperature dependence of the excitation energy due to interactions between magnons, librons and phonons. Phonon anharmonicities and the resulting expansion of the crystal lattice provide a simple way to view such effects. Our HET measurements probe the (011) Bragg peak for the $E_i = 60$ meV configuration. The determined $d$-spacing of this peak as a function of temperature is shown in Fig. 7(b) and agrees well with prior X-ray scattering measurement made at $T = 213(2)$ K[32]. The d-spacing of the (011) Bragg peak includes a contribution from the in-plane lattice constant distances of the $Cu_3$ trimer as shown in Fig. 1(a), and will be proportional to the temperature dependence of the distance between spins in the magnetic structure. Figure 7 clearly indicates that as the distance between interacting spins is increasing the exchange energy continuously decreases resulting in a softening of the magnetic excitation spectrum with increasing temperature.

**CONCLUSIONS**

The magnetic excitation spectrum in $Cu_3$ shows a considerable degree of overlap with the intramolecular vibrational spectrum. Thus, a straightforward interpretation of the $Cu_3$ scattering data in terms of well isolated magnetic excitations is difficult. However, we find consistent evidence that there is a magnetic excitation in $Cu_3$ at approximately 36.5 meV at low-temperatures. This was found via comparison with measurements of a non-magnetic HTiPB sample. In particular, examination of the wave-vector dependence of the $Cu_3$ data as well as the temperature dependence of the scattering intensity in the vicinity of the proposed magnetic excitation provided evidence for the presence of magnetic scattering. We determine a low-temperature value of the magnetic excitation in $Cu_3$ as being approximately 36.5(4) meV corresponding to an equilateral intratrimer exchange of $J = 24.3(3)$ meV. From examination of the temperature dependent elastic and inelastic scattering we are also able to correlate an expansion in the crystal lattice with a softening in the magnetic excitations as the temperature is increased. We note that this is a different value of intratrimer exchange compared to the value determined by temperature dependent magnetization. However, because of the presence of degenerate ground states in the $Cu_3$ trimer INS is able to provide a more unambiguous measure of the excitation energy despite the difficulties associated with the large hydrogen content of the material. Systematic magnetization measurements of $Cu_3$ and HTiPB samples as well as samples of $Cu_3$ with Zn substituted for the Cu sites may allow for a further refinement the intratrimer exchange constant based upon bulk measurements.

The energy scales of the magnetic excitation in the $Cu_3$ trimer and the temperature dependence of the vibrational and magnetic modes allow for a clear observation of the magnetism in the $Cu_3$ compound using inelastic neutron scattering techniques. Selective or total deuteration may allow for more systematic control of the vibrational modes in order to more clearly observe the magnetic excitations; however, our results show that neutron scattering measurements must be performed and analyzed carefully for materials which are prohibitively complicated and/or expensive to deuterate and which have a large number of proton sites for each magnetic spin, *e.g.* for $Cu_3$ there are 50.8 hydrogen sites per spin ½. Our results also provide motivation for explicit tuning of the non-magnetic portions of such systems. Integrating electron-phonon interactions (i.e. Jahn-Teller effects or magneto-elastic coupling) in such systems may provide further degrees of freedom in molecular design, such that properly accounting for lattice/vibrational excitations may provide additional tunability of chemical and magnetic properties of this class of magnetic materials. Finally, we note that there been a recent flurry of activity in EPR and magnetization studies of new clusters of Cu3[50,51], Cu4[52] and Cu5[53], among others. INS investigations of these and related materials are anticipated and should significantly benefit from the present work.

We acknowledge C. A. Murillo for helpful discussions and sample production, and C. D. Frost and M. D. Lumsden for helpful discussions during and after the measurements. Research sponsored by the Division of Materials Sciences and Engineering, Office of Basic Energy Sciences, U.S. Department of Energy, under contract DE-AC05-00OR22725 with Oak Ridge National Laboratory, managed and operated by UT-Battelle, LLC. NSD thanks NSF –DMR grant # 0506946 for financial support.

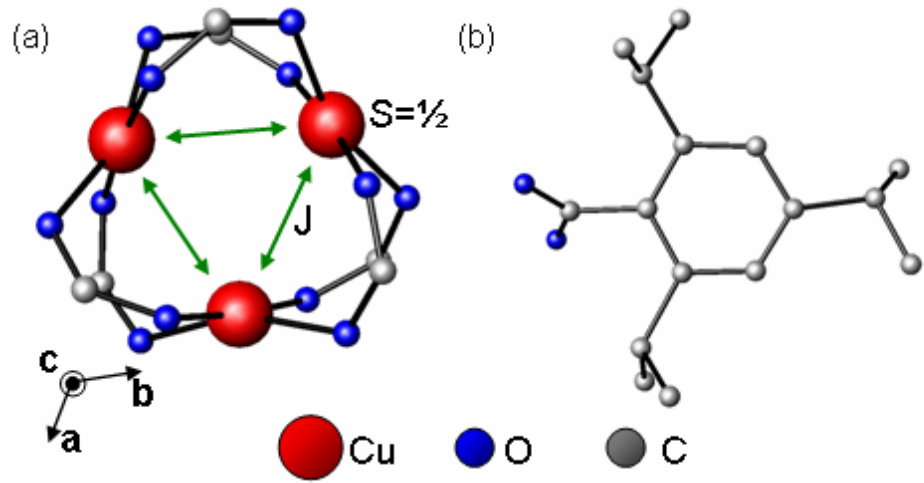

**Fig. 1** (color online) (a) Trimer structural unit of $Cu_3 \equiv Cu_3(O_2C_{16}H_{23})_6 \cdot 1.2C_6H_{12}$ as viewed along the c-axis of the trigonal crystal structure with $T = 213$ K lattice constants $a = b = 18.1331$ Å and $c = 19.4989$ Å [32]. Linear Cu-Cu distance is 3.131 Å. 2-4-6 triisopropyl-benzoic acid (HTiPB) carboxylate ligands attached to the oxygen sites are not shown. Arrows represent Cu-Cu spin exchange between $Cu^{2+}$ $S = 1/2$ sites, $J$. (b) Detail of oxygen sites with adjoining HTiPB carboxylate organic ligand as viewed normal to the ligand ring structure. A total of six of these units are attached to the trimer structure shown in (a). Hydrogen atomic sites are omitted for clarity. Panels (a) and (b) have different scales.

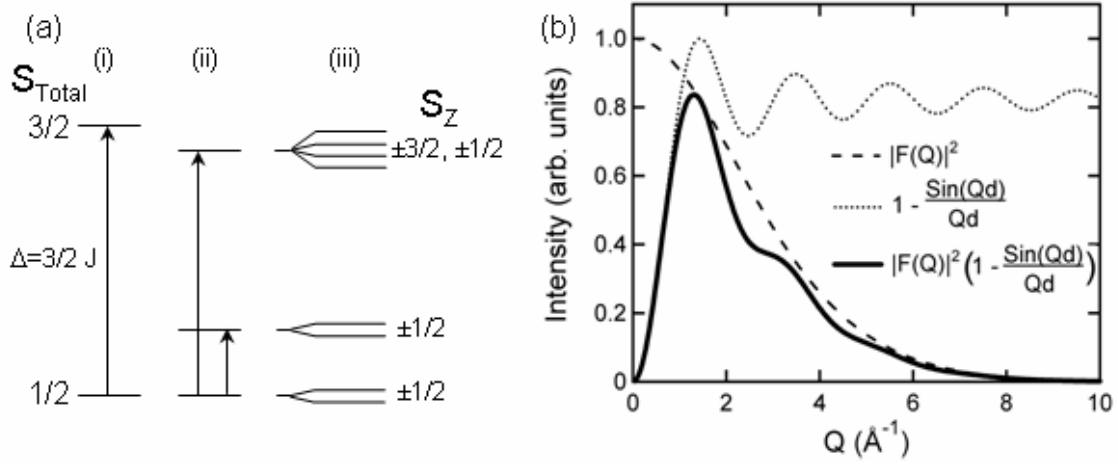

**Fig. 2** (a) Energy level diagrams of antiferromagnetic Heisenberg trimers. Case (i) equilateral trimer Heisenberg Hamiltonian $H = J(S_1 \cdot S_2 + S_2 \cdot S_3 + S_3 \cdot S_1)$ with a ground state consisting of a pair of degenerate $S_{Total} = 1/2$ doublets and an $S_{Total} = 3/2$ quartet excited state at an energy of $\Delta = 3/2J$. Case (ii) isosceles trimer with ground state degeneracy lifted due to asymmetric exchange. Case (iii) isosceles trimer in a finite magnetic field removes remaining degeneracy of ground and excited states via Zeman splitting. (b) Powder average equilateral trimer structure factor as described by Eq. 1. Dashed line represents the square of the $Cu^{2+}$ magnetic form factor, $|F(Q)|^2$ [54]. Dotted line corresponds to calculated powder average of the $S = 1/2$ trimer form factor for the Cu-Cu distances shown in Fig. 1(a). Solid line is the product of the dashed and dotted lines as described by Eq. 1.

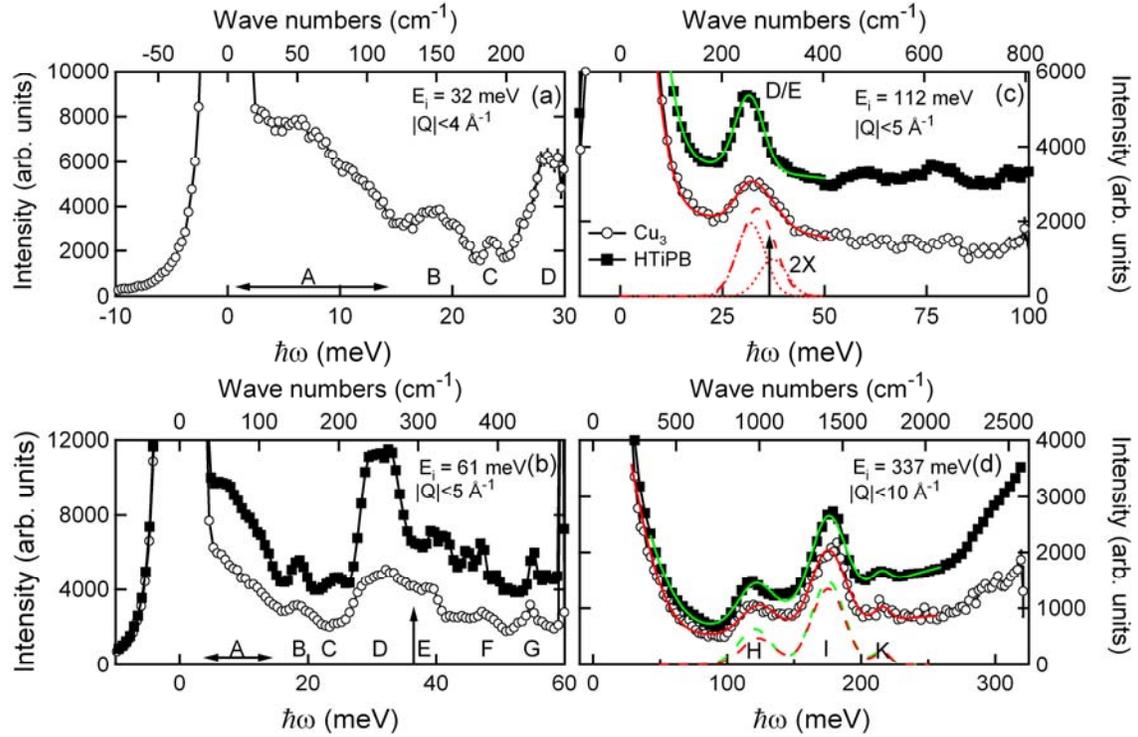

**Fig. 3** (color online) Measured INS intensity of $Cu_3$ (open symbols) and HTiPB (closed symbols) as a function of energy transfer. Labels *A* through *K* along the horizontal axes correspond to HTiPB vibrational modes discussed in the text. Data in (a), (b) and (d) were measured at $4 \leq T \leq 11$ K. Data in (c) were measured at $T = 80$ K. Vertical arrow in (b) and (c) represents location of proposed magnetic excitation. Bottom axes are plotted in units of meV and top axes are plotted in units of cm$^{-1}$. Solid green line in (b) is a $12 < \hbar\omega < 50$ meV fit to an inelastic Gaussian (FWHM = 8.070(2) meV) with a sloping background and an elastic Gaussian peak. Solid red line in (b) is a fit to two inelastic Gaussians with fixed FWHM = 8.070 meV and a sloping background and elastic Gaussian peak. Dashed[Dotted] line[s] are the single[two] $Cu_3$ fitted inelastic Gaussian peak[s] shown at 2X scale with no background terms. Solid lines in (d) are multiple Gaussian fits for $40 < \hbar\omega < 260$ meV as described in the text. Dashed lines are inelastic peaks from Gaussian fits in the absence of background terms.

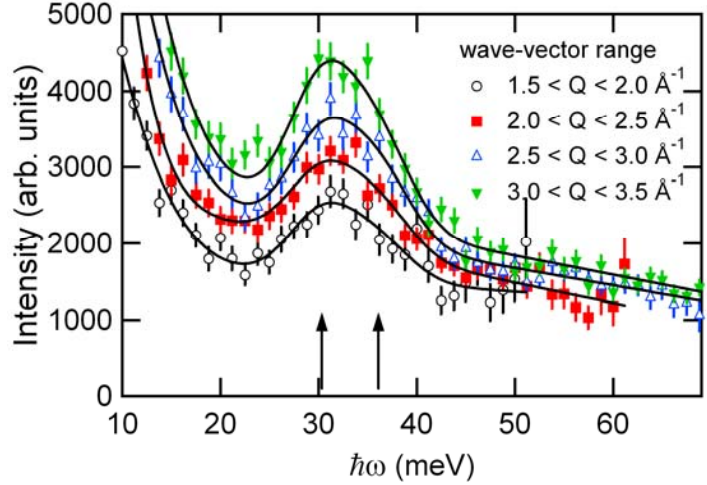

**Fig. 4** (color online) $T = 80$ K measured INS intensity of $Cu_3$ as a function of energy transfer integrated over different ranges of wave-vector transfer. Measurement performed with $E_i = 112$ meV. Solid lines are the result of a global fit to two fixed width Gaussians with a sloping background and elastic Gaussian peak as described in the text. Gaussian width is fixed to that found for vibrational modes in the non-magnetic sample in the same region of the measured spectrum as shown in Fig. 3(c). The peak locations of the two inelastic Gaussians, shown as vertical arrows, are a common fitting parameter and have values of 30.3(3) meV and 36.1(4) meV respectively.

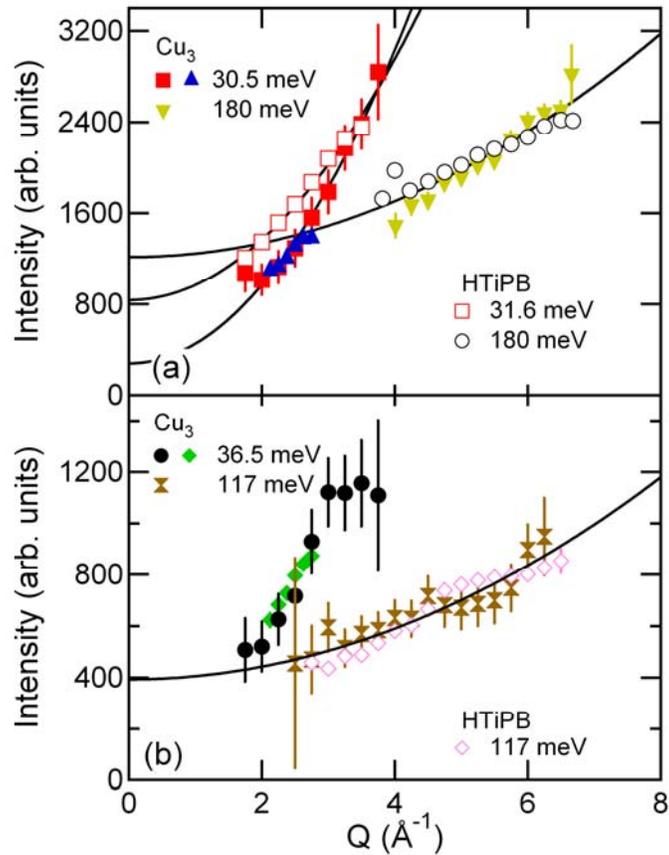

**Fig. 5** (color online) Fitted scattering intensity as a function of wave-vector transfer for inelastic excitations in Cu3 (closed symbols) and HTiPB (open symbols) in the vicinity of the respective energy transfers noted in the figure legend. Data have been scaled by multiplicative constants for presentation. Solid lines are quadratic fits with constant offsets as described in the text.

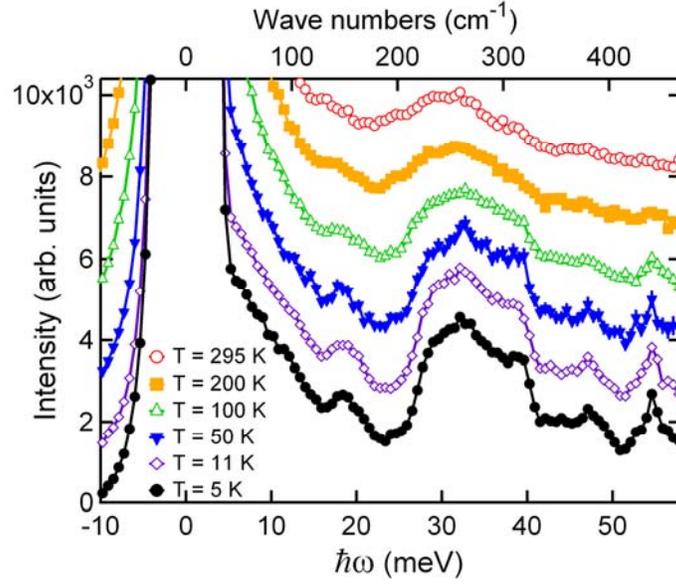

**Fig. 6** (color online) Measured INS intensity of the $Cu_3$ sample as a function of energy transfer integrated between $Q = 0$ and $Q = 5$ Å$^{-1}$ with $E_i = 61$ meV. Data for different temperatures have been offset along the vertical axis for presentation.

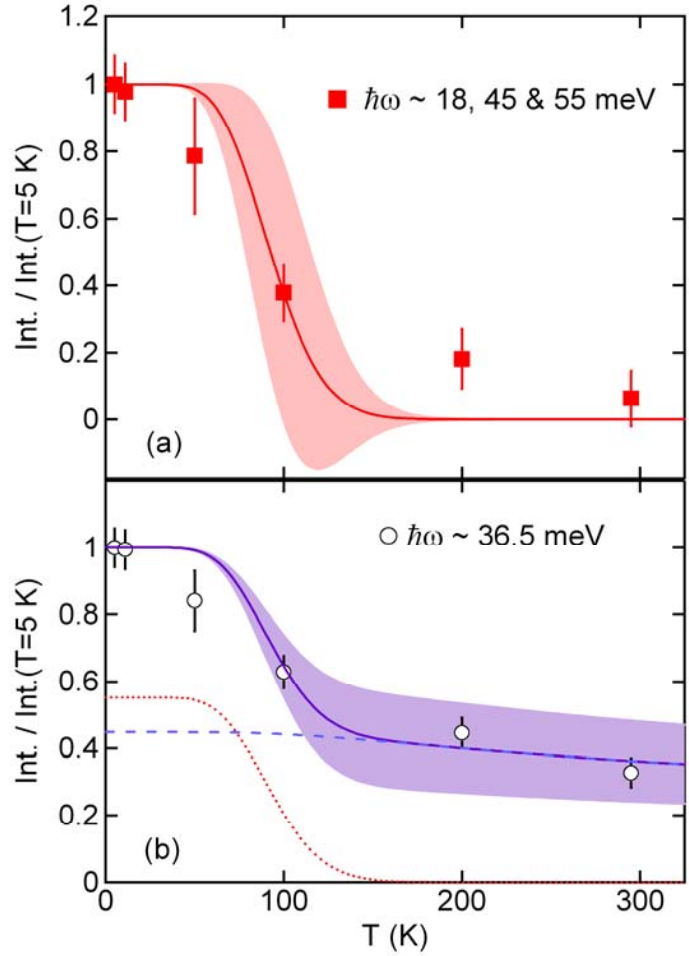

**Fig. 7** (color online) $Cu_3$ Temperature dependent normalized integrated scattering intensity associated with the (a) 18, 45 and 55 meV vibrational modes and (b) the integrated scattering intensity between 22 and 41 meV. Solid red line is single parameter fit to Eq(3) for the temperature dependence of molecular excitations. The blue solid line is a single parameter fit to the weighted sum of Eqs (2) and (3) using the Debye-Waller parameter found for the data in panel (a) as described in the text. Dashed blue and dotted red lines correspond to the respective magnetic and vibrational components of the fitted solid line in panel (b). Shaded regions correspond to the 2-sigma confidence bands of the single parameter fits.

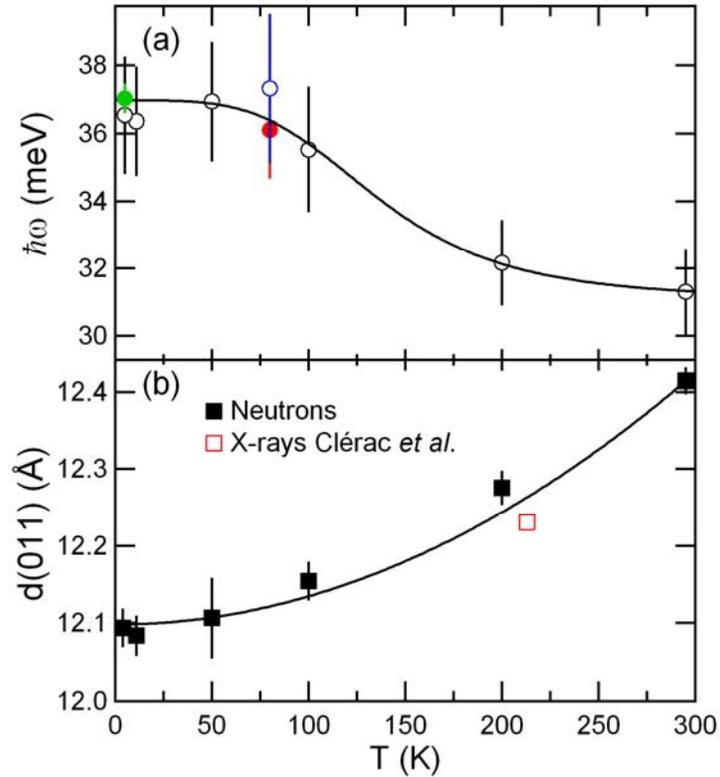

**Fig. 8** (color online) (a) Temperature dependent peak location of proposed magnetic excitation in $Cu_3$. Color (black) symbols correspond to peak locations determined in analysis of data in Figs. 3 and 4 (Fig. 6). Open (closed) symbols correspond to integrated (individual) wave-vector fits. Error bars are based upon the sum of the error in the fitted peak location and ten percent of the fitted FWHM. Solid line is a phenomenological fit as described in the text. (b) Temperature dependent d-spacing associated with the (011) Bragg peak as measured for $Cu_3$. A single data point is also plotted representing the result found in the prior X-ray scattering measurement[32]. Solid line is a polynomial fit including only a quadratic and constant term serving as a guide to the eye.